\RequirePackage{lineno}  

\documentclass[aps,prl,preprint,superscriptaddress,showpacs]{revtex4} 

\usepackage{graphicx}
\usepackage{rotating}
\usepackage{subfigure}
\usepackage{slashed}

\begin{document}
%
\newcommand{\fix}[1]{} 
\def\EtMiss {\slashed{E}_{T}}
\def\EtMissVec {\vec{\slashed{E}}_{T}}  
\def\DZero{D0~}
\setlength{\topmargin}{-15mm}



\title{Search for top-quark production via flavor-changing neutral currents in W+1 jet 
       events at CDF}  

\noaffiliation
\affiliation{Institute of Physics, Academia Sinica, Taipei, Taiwan 11529, Republic of China} 
\affiliation{Argonne National Laboratory, Argonne, Illinois 60439} 
\affiliation{University of Athens, 157 71 Athens, Greece} 
\affiliation{Institut de Fisica d'Altes Energies, Universitat Autonoma de Barcelona, E-08193, Bellaterra (Barcelona), Spain} 
\affiliation{Baylor University, Waco, Texas  76798} 
\affiliation{Istituto Nazionale di Fisica Nucleare Bologna, $^x$University of Bologna, I-40127 Bologna, Italy} 
\affiliation{Brandeis University, Waltham, Massachusetts 02254} 
\affiliation{University of California, Davis, Davis, California  95616} 
\affiliation{University of California, Los Angeles, Los Angeles, California  90024} 
\affiliation{University of California, San Diego, La Jolla, California  92093} 
\affiliation{University of California, Santa Barbara, Santa Barbara, California 93106} 
\affiliation{Instituto de Fisica de Cantabria, CSIC-University of Cantabria, 39005 Santander, Spain} 
\affiliation{Carnegie Mellon University, Pittsburgh, PA  15213} 
\affiliation{Enrico Fermi Institute, University of Chicago, Chicago, Illinois 60637}
\affiliation{Comenius University, 842 48 Bratislava, Slovakia; Institute of Experimental Physics, 040 01 Kosice, Slovakia} 
\affiliation{Joint Institute for Nuclear Research, RU-141980 Dubna, Russia} 
\affiliation{Duke University, Durham, North Carolina  27708} 
\affiliation{Fermi National Accelerator Laboratory, Batavia, Illinois 60510} 
\affiliation{University of Florida, Gainesville, Florida  32611} 
\affiliation{Laboratori Nazionali di Frascati, Istituto Nazionale di Fisica Nucleare, I-00044 Frascati, Italy} 
\affiliation{University of Geneva, CH-1211 Geneva 4, Switzerland} 
\affiliation{Glasgow University, Glasgow G12 8QQ, United Kingdom} 
\affiliation{Harvard University, Cambridge, Massachusetts 02138} 
\affiliation{Division of High Energy Physics, Department of Physics, University of Helsinki and Helsinki Institute of Physics, FIN-00014, Helsinki, Finland} 
\affiliation{University of Illinois, Urbana, Illinois 61801} 
\affiliation{The Johns Hopkins University, Baltimore, Maryland 21218} 
\affiliation{Institut f\"{u}r Experimentelle Kernphysik, Universit\"{a}t Karlsruhe, 76128 Karlsruhe, Germany} 
\affiliation{Center for High Energy Physics: Kyungpook National University, Daegu 702-701, Korea; Seoul National University, Seoul 151-742, Korea; Sungkyunkwan University, Suwon 440-746, Korea; Korea Institute of Science and Technology Information, Daejeon, 305-806, Korea; Chonnam National University, Gwangju, 500-757, Korea} 
\affiliation{Ernest Orlando Lawrence Berkeley National Laboratory, Berkeley, California 94720} 
\affiliation{University of Liverpool, Liverpool L69 7ZE, United Kingdom} 
\affiliation{University College London, London WC1E 6BT, United Kingdom} 
\affiliation{Centro de Investigaciones Energeticas Medioambientales y Tecnologicas, E-28040 Madrid, Spain} 
\affiliation{Massachusetts Institute of Technology, Cambridge, Massachusetts  02139} 
\affiliation{Institute of Particle Physics: McGill University, Montr\'{e}al, Qu\'{e}bec, Canada H3A~2T8; Simon Fraser University, Burnaby, British Columbia, Canada V5A~1S6; University of Toronto, Toronto, Ontario, Canada M5S~1A7; and TRIUMF, Vancouver, British Columbia, Canada V6T~2A3} 
\affiliation{University of Michigan, Ann Arbor, Michigan 48109} 
\affiliation{Michigan State University, East Lansing, Michigan  48824}
\affiliation{Institution for Theoretical and Experimental Physics, ITEP, Moscow 117259, Russia} 
\affiliation{University of New Mexico, Albuquerque, New Mexico 87131} 
\affiliation{Northwestern University, Evanston, Illinois  60208} 
\affiliation{The Ohio State University, Columbus, Ohio  43210} 
\affiliation{Okayama University, Okayama 700-8530, Japan} 
\affiliation{Osaka City University, Osaka 588, Japan} 
\affiliation{University of Oxford, Oxford OX1 3RH, United Kingdom} 
\affiliation{Istituto Nazionale di Fisica Nucleare, Sezione di Padova-Trento, $^y$University of Padova, I-35131 Padova, Italy} 
\affiliation{LPNHE, Universite Pierre et Marie Curie/IN2P3-CNRS, UMR7585, Paris, F-75252 France} 
\affiliation{University of Pennsylvania, Philadelphia, Pennsylvania 19104}
\affiliation{Istituto Nazionale di Fisica Nucleare Pisa, $^z$University of Pisa, $^{aa}$University of Siena and $^{bb}$Scuola Normale Superiore, I-56127 Pisa, Italy} 
\affiliation{University of Pittsburgh, Pittsburgh, Pennsylvania 15260} 
\affiliation{Purdue University, West Lafayette, Indiana 47907} 
\affiliation{University of Rochester, Rochester, New York 14627} 
\affiliation{The Rockefeller University, New York, New York 10021} 
\affiliation{Istituto Nazionale di Fisica Nucleare, Sezione di Roma 1, $^{cc}$Sapienza Universit\`{a} di Roma, I-00185 Roma, Italy} 

\affiliation{Rutgers University, Piscataway, New Jersey 08855} 
\affiliation{Texas A\&M University, College Station, Texas 77843} 
\affiliation{Istituto Nazionale di Fisica Nucleare Trieste/Udine, I-34100 Trieste, $^{dd}$University of Trieste/Udine, I-33100 Udine, Italy} 
\affiliation{University of Tsukuba, Tsukuba, Ibaraki 305, Japan} 
\affiliation{Tufts University, Medford, Massachusetts 02155} 
\affiliation{Waseda University, Tokyo 169, Japan} 
\affiliation{Wayne State University, Detroit, Michigan  48201} 
\affiliation{University of Wisconsin, Madison, Wisconsin 53706} 
\affiliation{Yale University, New Haven, Connecticut 06520} 
\author{T.~Aaltonen}
\affiliation{Division of High Energy Physics, Department of Physics, University of Helsinki and Helsinki Institute of Physics, FIN-00014, Helsinki, Finland}
\author{J.~Adelman}
\affiliation{Enrico Fermi Institute, University of Chicago, Chicago, Illinois 60637}
\author{T.~Akimoto}
\affiliation{University of Tsukuba, Tsukuba, Ibaraki 305, Japan}
\author{B.~\'{A}lvarez~Gonz\'{a}lez$^s$}
\affiliation{Instituto de Fisica de Cantabria, CSIC-University of Cantabria, 39005 Santander, Spain}
\author{S.~Amerio$^y$}
\affiliation{Istituto Nazionale di Fisica Nucleare, Sezione di Padova-Trento, $^y$University of Padova, I-35131 Padova, Italy} 

\author{D.~Amidei}
\affiliation{University of Michigan, Ann Arbor, Michigan 48109}
\author{A.~Anastassov}
\affiliation{Northwestern University, Evanston, Illinois  60208}
\author{A.~Annovi}
\affiliation{Laboratori Nazionali di Frascati, Istituto Nazionale di Fisica Nucleare, I-00044 Frascati, Italy}
\author{J.~Antos}
\affiliation{Comenius University, 842 48 Bratislava, Slovakia; Institute of Experimental Physics, 040 01 Kosice, Slovakia}
\author{G.~Apollinari}
\affiliation{Fermi National Accelerator Laboratory, Batavia, Illinois 60510}
% ...

\collaboration{CDF Collaboration\footnote{With visitors from $^a$University of Massachusetts Amherst, Amherst, Massachusetts 01003,
$^b$Universiteit Antwerpen, B-2610 Antwerp, Belgium, 
$^c$University of Bristol, Bristol BS8 1TL, United Kingdom,
$^d$Chinese Academy of Sciences, Beijing 100864, China, 
$^e$Istituto Nazionale di Fisica Nucleare, Sezione di Cagliari, 09042 Monserrato (Cagliari), Italy,
$^f$University of California Irvine, Irvine, CA  92697, 
$^g$University of California Santa Cruz, Santa Cruz, CA  95064, 
$^h$Cornell University, Ithaca, NY  14853, 
$^i$University of Cyprus, Nicosia CY-1678, Cyprus, 
$^j$University College Dublin, Dublin 4, Ireland,
$^k$University of Edinburgh, Edinburgh EH9 3JZ, United Kingdom, 
$^l$University of Fukui, Fukui City, Fukui Prefecture, Japan 910-0017
$^m$Kinki University, Higashi-Osaka City, Japan 577-8502
$^n$Universidad Iberoamericana, Mexico D.F., Mexico,
$^o$Queen Mary, University of London, London, E1 4NS, England,
$^p$University of Manchester, Manchester M13 9PL, England, 
$^q$Nagasaki Institute of Applied Science, Nagasaki, Japan, 
$^r$University of Notre Dame, Notre Dame, IN 46556,
$^s$University de Oviedo, E-33007 Oviedo, Spain, 
$^t$Texas Tech University, Lubbock, TX  79609, 
$^u$IFIC(CSIC-Universitat de Valencia), 46071 Valencia, Spain,
$^v$University of Virginia, Charlottesville, VA  22904,
$^w$Bergische Universit\"at Wuppertal, 42097 Wuppertal, Germany,
$^{ee}$On leave from J.~Stefan Institute, Ljubljana, Slovenia, 
}}
\noaffiliation

\date{\today}

\begin{abstract}
We report on a search for the non-standard-model process $u(c)+g\rightarrow t$ 
using $p{\bar{p}}$ collision data collected by the CDF II detector 
corresponding to $2.2\,\mathrm{fb^{-1}}$. 
The candidate events are classified as signal-like or background-like by an 
artificial neural network.
The observed discriminant distribution yields no evidence for FCNC top-quark 
production, resulting in an upper limit on the production cross section 
$\sigma (u(c)+g\rightarrow t) < 1.8\;\mathrm{pb}$ at the 95\% C.L.
Using theoretical predictions we convert the 
cross-section limit to upper limits 
on FCNC branching ratios:
$\mathcal{B}(t\rightarrow u+g) < 3.9 \times 10^{-4}$ and
$\mathcal{B}(t\rightarrow c+g) < 5.7 \times 10^{-3}$.
\end{abstract}



\maketitle

In the standard model (SM) of particle physics the flavor quantum number of fermions 
can be changed by charged currents, i.e., weak interactions mediated by the exchange 
of a $W^\pm$ boson. Flavor-changing neutral-currents (FCNC) are absent at tree level, 
but do occur at higher order in perturbation theory
through loop diagrams. These radiative corrections are further suppressed through
the GIM mechanism~\cite{Glashow:1970gm}.
In the bottom-quark sector the large top-quark mass alleviates the GIM suppression
leading to FCNC decays with branching ratios at the level of $10^{-6}$, while
in the top-quark sector FCNC decays are more strongly suppressed and occur only at
the order of $\mathcal{B}\approx 10^{-10}$ to $10^{-14}$~\cite{eilam1991},
way beyond the current experimental sensitivity.
Therefore, any evidence for FCNC in the top-quark sector will be a signal
of physics beyond the SM. Enhanced FCNC effects can be realized in extensions
of the SM, such as models with multiple Higgs 
doublets~\cite{eilam1991,Hou1992}, supersymmetric models with
R-parity violation~\cite{yang1998}, or topcolor-assisted technicolor 
theories~\cite{yue2001}. In certain regions of parameter space of these models
the branching ratio of FCNC decays can reach levels of $10^{-3}$ to $10^{-5}$. But even with
such an enhancement the detection of FCNC top-quark decays remains a very 
challenging task at the Tevatron: First, because one can only expect to
reconstruct a few top quarks in these modes, and second, because the background
for the most promising mode, $t\rightarrow cg$, is very difficult to discern
from generic multijet production via quantum chromodynamics (QCD).  
It has therefore been suggested to search for FCNC couplings in top-quark 
production, rather than top-quark 
decay~\cite{hosch1997,tait2000}.

In this Letter we present a search for the non-SM single top-quark production
processes $u(c)+g \rightarrow t$. We do not consider a particular model, but perform
a model-independent search based on an effective theory~\cite{hosch1997} 
that contains
additional flavor-changing operators in the Lagrangian
\begin{equation}
  \label{eq:fcnc}
  g_s \frac{\kappa_{tug}}{\Lambda} \;\bar{u}\;\sigma^{\mu\nu} \frac{\lambda^a}{2}
  t\; G^{a}_{\mu\nu} + 
  g_s \frac{\kappa_{tcg}}{\Lambda} \;\bar{c}\;\sigma^{\mu\nu} \frac{\lambda^a}{2}
  t\; G^{a}_{\mu\nu} +   
  \ h.c.
\end{equation}
Here $\kappa_{tug}$ and $\kappa_{tcg}$ are dimensionless parameters that relate 
the strength of 
the new, anomalous coupling to the strong coupling constant $g_s$ and
$\Lambda$ is the new physics scale, related to the mass cutoff above which the
effective theory breaks down. The gluon field tensor is denoted $G^a_{\mu\nu}$,
the $\lambda^a$ are the Gell-Mann matrices, and
$\sigma^{\mu\nu} \equiv \frac{i}{2}[\gamma^\mu, \gamma^\nu]$ transforms as a tensor under
the Lorentz group. 
The existence of FCNC operators
allows the production of top quarks via $u(c)+g \rightarrow t$, but also
non-SM decays $t\rightarrow u(c)+g$. 
In the allowed region of parameter space for $\kappa_{tug}$ and $\kappa_{tcg}$ an 
experimentally favorable situation occurs. While the FCNC production 
cross-section of top quarks is in the range of several picobarns, 
the branching ratio of FCNC decays is very small, and top quarks can thus be 
reconstructed in the SM decay mode $t\rightarrow Wb$.
While $u$ quarks are constituent quarks of the proton, $c$ quarks, as needed for the
process $c+g \rightarrow t$, occur as
sea quarks originating from a gluon splitting into a $c\bar{c}$ pair.
In the SM, top quarks are either produced as $t\bar{t}$ pairs by the strong
interaction or singly via the exchange of a virtual $W$ boson. The pair-production
process is firmly established experimentally with a cross section of about 
$7\,\mathrm{pb}$. Evidence for SM single top-quark production has been shown by
CDF~\cite{CDFst} and D\O~\cite{D0st}, yielding a cross section around $3\,\mathrm{pb}$.

Our analysis is the first one at the Tevatron searching for the $2\rightarrow 1$
processes $u(c)+g \rightarrow t$, while a previous D\O \ analysis~\cite{fcncd0} 
has looked for $2\rightarrow 2$ processes, such as
$q\bar{q}\rightarrow t\bar{u}$, $ug\rightarrow tg$, or $gg\rightarrow t\bar{u}$,
resulting in upper limits 
of $\kappa_{tug}/\Lambda < 0.037\;\mathrm{TeV^{-1}}$ and
$\kappa_{tcg}/\Lambda < 0.15\;\mathrm{TeV^{-1}}$ at the 95\% C.L.  
FCNC couplings to the top quark involving the photon or $Z$ boson
have been constrained by the analysis of top-quark decays at the 
Tevatron~\cite{CDFfcncDecay}, the search for 
$e^+e^- \rightarrow t\bar{c}/t\bar{u}$ reactions at LEP, see e.g.~\cite{L3fcnc}, 
and the search for $ep\rightarrow e+t+X$ reactions at HERA~\cite{fcncZEUS,fcncH1}.

The analysis presented here uses $p\bar{p}$ collision data at 
$\sqrt{s}=1.96\;\mathrm{TeV}$ collected by the CDF II detector~\cite{CDFdet} at 
the Fermilab Tevatron between March 2002 and August 2007.
The data set corresponds to an integrated luminosity of $2.2\;\mathrm{fb^{-1}}$.
We select a set of candidate events in the $t\rightarrow Wb\rightarrow \ell\nu b$ 
topology based on the event selection used for the measurement of SM 
single top-quark production~\cite{CDFst}. 
We require exactly one isolated~\cite{isolation} electron with 
transverse energy~\cite{etadef}
$E_T>20\,\mathrm{GeV}$ or one isolated muon with 
$p_T > 20\,\mathrm{GeV}/c$, 
missing transverse energy $\EtMiss > 25\;\mathrm{GeV}$,
and exactly one jet with pseudorapidity~\cite{etadef} $|\eta|\leq 2.8$ and 
$E_T >20\,\rm{GeV}$.
The jet is further required to
contain a reconstructed secondary vertex consistent with the decay of a
$b$ hadron~\cite{secvtx}. 
After all selection cuts we observe 2472 candidate events. 

Background yields from diboson processes $WW$, $WZ$, $ZZ$, and $t{\bar{t}}$ production 
are predicted using {\sc pythia}~\cite{pythia}
Monte Carlo samples, normalized to next-to-leading order (NLO) 
cross sections~\cite{diboson,ttbarxs}.
SM single top-quark rates are estimated with simulated events from 
the tree-level matrix-element generator {\sc madevent}~\cite{madevent},
subsequent showering with {\sc pythia}, and normalization to NLO cross
sections~\cite{stxs}.
The processes with vector bosons ($W$ or $Z$) plus jets are generated with 
{\sc alpgen}~\cite{alpgen}, with parton showering and underlying event simulated 
with {\sc pythia}.
Using a compound model~\cite{CDFst} based on simulated events, 
theoretical cross sections, and 
normalizations in background-dominated regions we predict the composition of the $W$+1 jet data
set as given in Table~\ref{tab:eventyield}.
%
\begin{table}[t!]
\caption{Predicted sample composition and observed number of 
  $W$+1 jet events in 2.2~fb$^{-1}$ of CDF Run II data.}
\begin{center}
\begin{tabular}{lrcr}
\hline
\hline
Process & \multicolumn{3}{c}{Expected events}\\
\hline
$Wb\bar{b}$, $Wc\bar{c}$ & 750.9 & $\pm$ & 225.3 \\
$Wc$                    & 622.3 & $\pm$ & 186.7 \\
$Wq\bar{q}$             & 769.9 & $\pm$ & 100.5 \\
$t\bar{t}$              & 12.3 & $\pm$ & 1.8 \\
QCD-multijet            & 43.0 & $\pm$ & 17.2 \\
Diboson                 & 19.9 & $\pm$ & 2.0 \\
$Z$+jets                & 26.6 & $\pm$ & 4.2 \\
SM single-top           & 24.4 & $\pm$ & 3.6 \\
\hline
Total prediction $\ \ $ & 2269.3 & $\pm$ & 434.3 \\
\hline
Observed                & \multicolumn{3}{c}{2472}               \\
\hline
\hline
\end{tabular}
\end{center}
\label{tab:eventyield}
\end{table}
Top-quark events produced via the processes $u(c)+g\rightarrow t$ are simulated using the
matrix-element generator {\sc toprex}~\cite{toprex}, followed by parton showering with
{\sc pythia}. 
For the event generation, the coupling constants have been chosen to yield
a cross section of $1\,\mathrm{pb}$, 
which correponds to the approximate sensitivity to the process with the data set we analyzed.
By investigating kinematic distributions at parton level, we verified that the event 
kinematics do not depend on that choice of parameters within the range relevant for our
analysis.  
Under the assumption that $\kappa_{tug}=\kappa_{tcg}$ the $tug$ coupling contributes 
$0.94\,\mathrm{pb}$ and the $tcg$ coupling $0.06\,\mathrm{pb}$. 
For a total FCNC top-quark cross section of $1\,\mathrm{pb}$ we expect a yield
of $35.3\pm 5.3$ events. 

For an efficient background rejection, we employ the same neural-network technology
as used in the search for SM single top-quark production~\cite{CDFst,Richter:2007zzc}. 
Neural networks (NN) have the advantage that correlations between the discriminating 
input variables are identified and utilized to optimize the separation power between
signal and background processes. The networks are developed using the {\sc neurobayes} analysis 
package~\cite{Feindt:2006pm}, which combines a three-layer feed-forward neural network with 
a complex and robust preprocessing of the input variables. 
The network infrastructure consists of one input node for each input variable
plus one bias node, 
15 hidden nodes, and one output node,
which gives a continuous output in the interval [$-1,1$]. We train the NN on the samples of 
simulated events listed above using a mixture of 50\% signal events and 50\% 
background events. The background composition is chosen in the proportions given in
Table~\ref{tab:eventyield}, with SM single top-quark events included as background. 
In total, we use 14 variables that show significant discriminating power 
between signal and background.
Variables derived directly from the four-vectors of reconstructed particles are 
the $p_T$ and the $\eta$ of 
the charged lepton, the $p_T$ of the jet, the difference in azimuth 
angle between the jet and $\EtMissVec$, and between the lepton and $\EtMissVec$, as well as
the $\Delta R$ between the charged lepton and the jet.
The $W$-boson candidate is reconstructed in its leptonic decay mode from the charged lepton
and $\EtMissVec$ applying the kinematical constraint 
$M_{\ell\nu} = M_W = 80.4\,\mathrm{GeV}/c^2$. The two-fold
ambiguity for the $z$-component of the neutrino momentum is resolved by choosing the
smaller $|p_{z,\nu}|$ solution. Based on the $W$-boson reconstruction we define two input 
variables: the transverse mass $M_{T,\ell\nu}$ and $\eta_{\ell\nu}$.
We further reconstruct top-quark candidates by adding the jet to the 
reconstructed $W$ boson and thereby define the following input variables:
$M_{\ell\nu j}$, $M_{T,\ell\nu j}$, the rapidity $y_{\ell\nu j}$, 
and $Q_\ell\cdot\eta_{\ell\nu j}$ where $Q_\ell$ is the charge of the lepton. 
An additional input variable is the output of an advanced jet-flavor separating 
tool mainly developed to increase the sensitivity of the SM single top-quark
searches~\cite{Richter:2007zzc}. 
To describe the event shape in general, we use the aplanarity of the reconstructed 
top-quark decay system~\cite{aplanarity}. 

\begin{figure}[t]
\subfigure[]{
  \label{fig:NNoutShapes}
  \includegraphics[width=0.7\columnwidth]{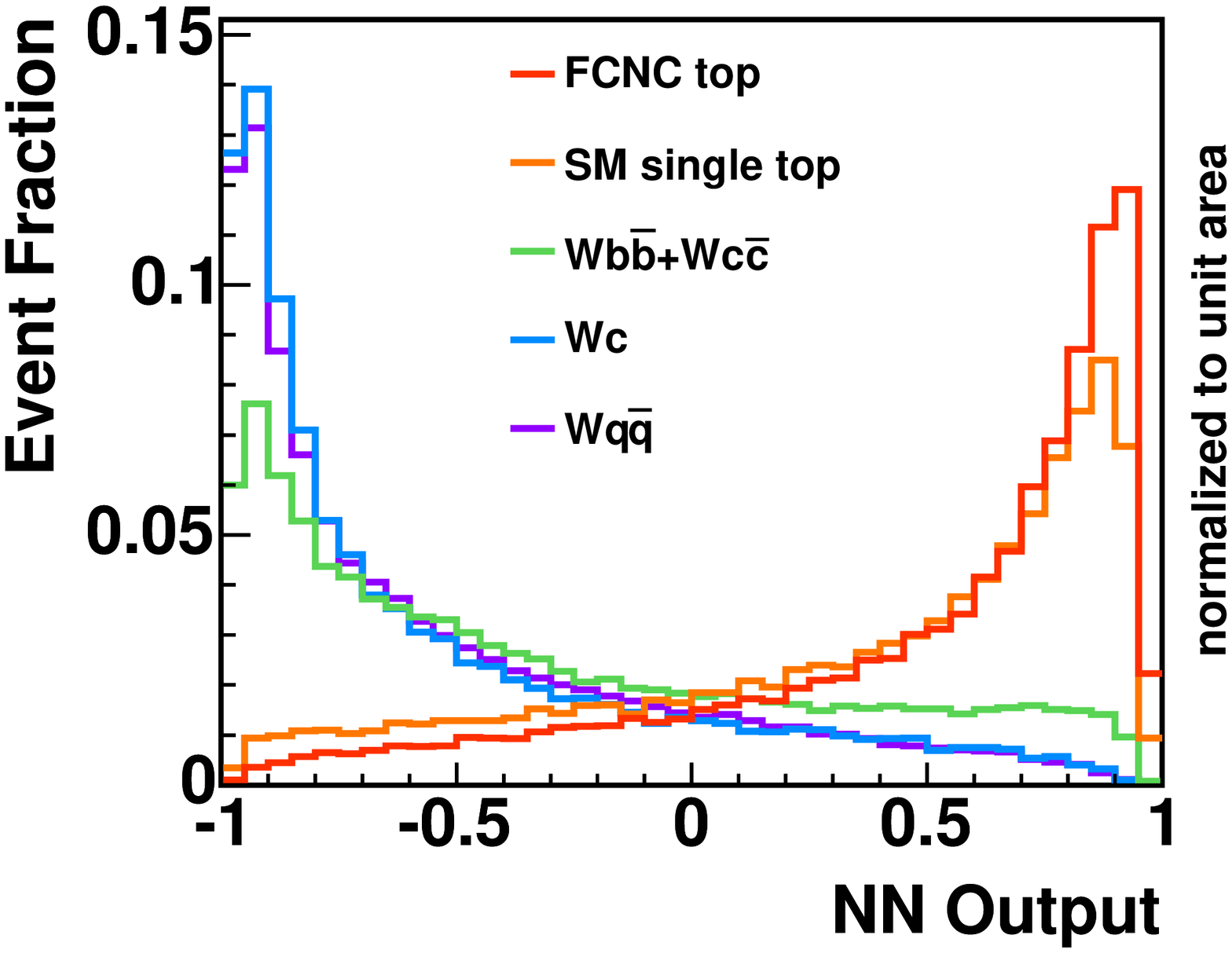}
}
\subfigure[]{
  \label{fig:NNoutStack}
  \includegraphics[width=0.7\columnwidth]{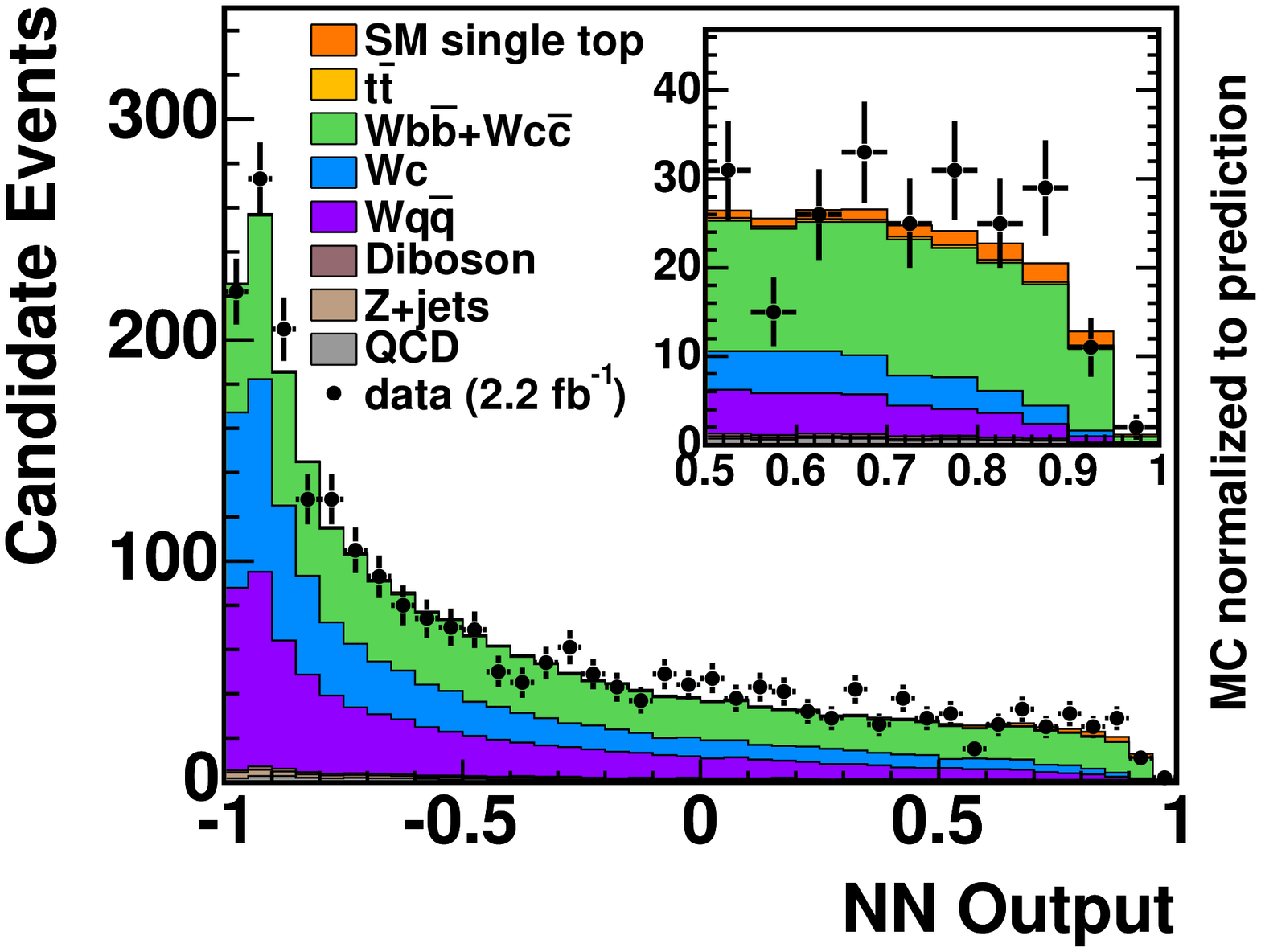}
}
\caption{Distribution of the NN discriminant. \subref{fig:NNoutShapes} Discriminant shapes for the
  different physics processes normalized to unit area. 
  \subref{fig:NNoutStack} The composite model is compared to the
  distribution observed in collision data. The inset shows the high NN-output region, 
  where top-quark events contribute the most.}
\label{fig:nnout}
\end{figure}

We apply the NN to the samples of simulated events and obtain template distributions 
of the network output for all physics processes considered. 
The template distributions of the most important background processes and the 
signal are shown in Fig.~\ref{fig:nnout}\subref{fig:NNoutShapes}.  
As can be seen, the separation between FCNC top-quark events and SM single top-quark
events is only marginal.
The templates are 
weighted by their expected event yields and the resulting composite model is compared
to the NN output distribution observed in collision data in 
Fig.~\ref{fig:nnout}\subref{fig:NNoutStack}.
%

To measure the potential content of FCNC-produced top quarks in the observed
data set, we perform a binned maximum likelihood fit of the NN output distribution.
The effect of systematic uncertainties is parameterized in the likelihood function
including the correlation of rate normalization effects and shape distortions of the template 
distributions.
Uncertainties in the jet energy scale, $b$-tagging efficiencies, lepton identification
and trigger efficiencies, the amount of initial and final
state radiation, parton distribution functions, factorization and renormalization scale dependence, 
and Monte Carlo modeling have been explored and incorporated in this analysis.
We integrate over all parameters describing systematic uncertainties in the likelihood 
function using Gaussian priors.
The rate of $Wb\bar{b}$ and $Wc\bar{c}$ events is required to be positive, but
otherwise unconstrained.
Applying a prior probability density, that is zero if the FCNC cross section is negative and one elsewhere, 
we obtain the posterior probability density. 
No significant rate of top quarks produced by FCNC is observed
and we set an upper limit on the cross section of $1.8\,\mathrm{pb}$ at the
95\% C.L., which is in good agreement with the expected upper limit of $1.3\,\mathrm{pb}$
obtained from ensemble tests. The probability to obtain an upper limit higher 
than the observed $1.8\,\mathrm{pb}$ under the assumption that FCNC top-quark
production does not exist is 28\%.  

Using theoretical predictions of $\sigma (u(c)+g\rightarrow t)$, 
which include threshold resummation
effects~\cite{Liu2006,Liu:2005dp}, we 
convert the upper limit on the cross section into upper limits on the FCNC 
coupling constants at the 95\% C.L. and find
$\kappa_{tug}/\Lambda < 0.018\;\mathrm{TeV^{-1}}$ assuming $\kappa_{tcg}=0$, and
$\kappa_{tcg}/\Lambda < 0.069\;\mathrm{TeV^{-1}}$ assuming $\kappa_{tug}=0$.
Using predictions at NLO~\cite{Liu2008}, we also express these limits on the
coupling constants in terms of limits on the FCNC branching ratios and obtain:
$\mathcal{B}(t\rightarrow u+g) < 3.9 \times 10^{-4}$ and
$\mathcal{B}(t\rightarrow c+g) < 5.7 \times 10^{-3}$.

For the first time we have explored the $W$+1 jet data set 
in search for top quarks produced by gluon-induced FCNC via the processes 
$u(c)+g\rightarrow t$. No evidence for such processes is found, resulting in 
the most stringent limits on the branching fractions for FCNC top-quark decays.

\begin{acknowledgments}
The authors express their gratitude to Chong Sheng Li of Peking University 
for very useful communication
and for providing a new calculation of FCNC top-quark branching ratios in a very
timely fashion.
We thank the Fermilab staff and the technical staffs of the participating institutions for their vital contributions. This work was supported by the U.S. Department of Energy and National Science Foundation; the Italian Istituto Nazionale di Fisica Nucleare; the Ministry of Education, Culture, Sports, Science and Technology of Japan; the Natural Sciences and Engineering Research Council of Canada; the National Science Council of the Republic of China; the Swiss National Science Foundation; the A.P. Sloan Foundation; the Bundesministerium f\"ur Bildung und Forschung, Germany; the Korean Science and Engineering Foundation and the Korean Research Foundation; the Science and Technology Facilities Council and the Royal Society, UK; the Institut National de Physique Nucleaire et Physique des Particules/CNRS; the Russian Foundation for Basic Research; the Ministerio de Ciencia e Innovaci\'{o}n, and Programa Consolider-Ingenio 2010, Spain; the Slovak R\&D Agency; and the Academy of Finland.
\end{acknowledgments}


\begin{thebibliography}{99}

\bibitem{Glashow:1970gm}S.L. Glashow, J. Iliopoulos, and L. Maiani,
  Phys. Rev. D {\bf 2}, 1285 (1970).

\bibitem{eilam1991}G. Eilam, J.L. Hewett, and A. Soni,
  Phys. Rev. D {\bf 44}, 1473 (1991); 
  Erratum: Phys. Rev. D {\bf 59}, 039901(E) (1998).


\bibitem{Hou1992}W.-S. Hou, Phys. Lett. B {\bf 296}, 179 (1992).


\bibitem{yang1998}J.M. Yang, B.-L. Young, and X. Zhang, Phys. Rev. D {\bf 58},
  055001 (1998).

\bibitem{yue2001}C. Yue {\it et al.}, Phys. Lett. B {\bf 508}, 290 (2001).


\bibitem{hosch1997} M. Hosch, K. Whisnant, and B.-L. Young, Phys. Rev. D
  {\bf 56}, 5725 (1997).



\bibitem{tait2000}T.M.P. Tait and C.-P. Yuan, Phys. Rev. D {\bf 63}, 014018 (2000).

\bibitem{CDFst}T. Aaltonen {\it et al.} (CDF Collaboration), arXiv:0809.2581 [hep-ex],
  accepted by Phys. Rev. Lett.

\bibitem{D0st}V.M. Abazov {\it et al.} (D\O \ Collaboration), 
  Phys. Rev. Lett. {\bf 98}, 181802 (2007).

\bibitem{fcncd0}V.M. Abazov {\it et al.} (D\O \ Collaboration), 
  Phys. Rev. Lett. {\bf 99}, 191802 (2007). 

\bibitem{CDFfcncDecay}T. Aaltonen {\it et al.} (CDF Collaboration), 
  Phys. Rev. Lett. {\bf 101}, 192002 (2008).

\bibitem{L3fcnc}P. Achard {\it et al.} (L3 Collaboration), Phys. Lett. B {\bf 549}, 
  290 (2002).

\bibitem{fcncZEUS}S. Chekanov {\it et al.} (ZEUS Collaboration), Phys. Lett. B {\bf 559}, 
  153 (2003).

\bibitem{fcncH1}A. Aktas {\it et al.} (H1 Collaboration), Eur. Phys. J. C {\bf 33}, 
  9 (2004). 

\bibitem{CDFdet}D. Acosta {\it et al.} (CDF Collaboration), Phys. Rev. D {\bf 71}, 
  032001 (2005).

\bibitem{etadef}In the CDF geometry, $\theta$ is the polar angle with respect to the proton beam 
axis, and $\phi$ is the azimuthal angle. Pseudorapidity is 
defined by $\eta \equiv -\ln[\tan(\theta/2)]$.
The transverse momentum $p_T$ is the component of the momentum projected 
onto the plane transverse 
to the beam axis. The transverse energy of a shower or calorimeter tower is 
defined as $E_T \equiv E\sin\theta$, 
where $E$ is the energy deposited. 

\bibitem{jesnim}A. Bhatti {\it et al.}, Nucl. Instrum. Methods. A {\bf 566}, 2 (2006).

\bibitem{isolation}D. Acosta {\it et al.} (CDF Collaboration), 
  Phys. Rev. D {\bf 72}, 052003 (2005).

\bibitem{secvtx}D. Acosta {\it et al.} (CDF Collaboration), 
  Phys. Rev. D {\bf 71}, 052003 (2005).

\bibitem{TPeiffer}T. Peiffer, diploma thesis, University of Karlsruhe, 2008,
  FERMILAB-MASTERS-2008-01.

\bibitem{pythia} T.~Sj\"ostrand {\it et al.}, Comput. Phys. Commun. {\bf 135}, 238 (2001).

\bibitem{diboson} J.M. Campbell and R.K. Ellis, 
  Phys.\ Rev. D {\bf 60}, 113006 (1999).                                       

\bibitem{ttbarxs}M. Cacciari {\it et al.}, J.\ High\ Energy\ Phys. {\bf 0404}, 068 (2004). 

\bibitem{madevent} J.~Alwall {\it et al.}, J.\ High\ Energy\ Phys. {\bf 0709}, 028 (2007).

\bibitem{stxs} B.W.~Harris, E. Laenen, L. Phaf, Z. Sullivan, S. Weinzierl, 
  Phys.\ Rev. D {\bf 66}, 054024 (2002).   

\bibitem{alpgen} M.L.~Mangano {\it et al.}, J.\ High\ Energy\ Phys. {\bf 0307}, 001 (2003).


\bibitem{toprex}S.R. Slabospitsky and L. Sonnenschein, Comput. Phys. Commun. {\bf 148},
  87 (2002).

\bibitem{Richter:2007zzc} S. Richter, Ph.D. thesis, University of Karlsruhe, 2007,
 FERMILAB-THESIS-2007-35.

\bibitem{Feindt:2006pm} M. Feindt and U. Kerzel,
  Nucl. Instrum. Methods A {\bf 559}, 190-194 (2006).

\bibitem{aplanarity}Aplanarity is defined as $\frac{3}{2}$ of the smallest eigenvalue 
  of the momentum tensor constructed from the jet, the charged lepton, and the 
  reconstructed neutrino.

\bibitem{Liu2006}
L.L. Yang, C.S. Li, Y. Gao, and J.J. Liu, 
  Phys. Rev. D {\bf 73}, 074017 (2006).
\bibitem{Liu:2005dp} 
  J.J. Liu, C.S. Li, L.L. Yang, and L.G. Jin,
  Phys. Rev. D {\bf 72}, 074018 (2005). 

\bibitem{Liu2008} J.J. Zhang {\it et al.}, Phys. Rev. Lett. {\bf 102}, 072001 (2009). 




\end{thebibliography}
\end{document}